\documentclass[12pt]{article}
\usepackage{amsmath,amssymb,amsfonts,amscd,bbm,graphicx,cite}
\begin{document}


\thispagestyle{empty}
\renewcommand{\thefootnote}{\fnsymbol{footnote}}
\setcounter{footnote}{1}

\vspace*{-1.cm}
\begin{flushright}
OSU-HEP-04-06
\end{flushright}
\vspace*{1.8cm}

\centerline{\Large\bf 6D Higgsless Standard Model}

\vspace*{18mm}

\centerline{\large\bf 
S. Gabriel\footnote{E-mail: \texttt{svengab@susygut.phy.okstate.edu}},
S. Nandi\footnote{E-mail: \texttt{shaown@okstate.edu}},
and 
G. Seidl\footnote{E-mail: \texttt{gseidl@susygut.phy.okstate.edu}}}
      
\vspace*{5mm}
\begin{center}
{\em Department of Physics, Oklahoma State University,}\\
{\em Stillwater, OK 74078, USA}
\end{center}

\vspace*{20mm}

\centerline{\bf Abstract}
\vspace*{2mm}
We present a six-dimensional Higgsless Standard Model with a realistic gauge
sector. The model uses only the Standard Model gauge group
$SU(2)_L\times U(1)_Y$ with the gauge bosons propagating in flat extra
dimensions compactified on a rectangle. The electroweak symmetry is broken by boundary conditions, and the correct splitting between the $W$ and $Z$ gauge
boson masses can be arranged by suitable choice of the compactification scales. The higher Kaluza-Klein excitations
of the gauge bosons decouple from the effective low-energy theory due to dominant brane kinetic terms. The model has the 
following two key features compared to five-dimensional models. The
dimensional couplings in the bulk Lagrangian, responsible for electroweak symmetry breaking using mixed boundary conditions, are of order the electroweak scale. Moreover, with respect to ``oblique'' corrections, the agreement with the precision electroweak parameters is improved compared to five-dimensional
warped or flat space models. We also argue that the calculability of
Higgsless models can be ameliorated in more than five dimensions.


\renewcommand{\thefootnote}{\arabic{footnote}}
\setcounter{footnote}{0}

\newpage

\section{Introduction}
The Standard Model (SM) of electroweak interactions \cite{Weinberg:tq}, based
on the gauge symmetry group $SU(2)_L\times U(1)_Y$, provides a highly successful
description of electroweak precision tests (EWPT) \cite{Hagiwara:fs,LEP}. One
fundamental ingredient of the SM is the Higgs mechanism \cite{Higgs:ia}, which accomplishes electroweak symmetry breaking (EWSB) and at high
energies unitarizes massive $W^\pm$ and $Z$ scattering through the presence of the
scalar Higgs doublet \cite{LlewellynSmith:ey}. However, no fundamental
scalar particle has been observed yet in Nature, and as long as there is no
direct evidence for the existence of the Higgs boson, the actual
mechanism of EWSB remains a mystery. In case the Higgs boson
will also not be found at the Tevatron or the LHC, it will therefore be
necessary to consider alternative ways to achieve EWSB without a Higgs.

It is well known, that in extra dimensions gauge symmetries can also be
broken by boundary conditions (BC's) on a compact space
\cite{Scherk:1978ta}. Here, a geometric
``Higgs'' mechanism ensures tree-level unitarity of longitudinal gauge boson
scattering through a tower of Kaluza-Klein (KK) \cite{Kaluza:tu} excitations
\cite{SekharChivukula:2001hz}. The SM in $(\rm TeV)^{-1}$-sized extra dimensions with gauge
symmetry breaking by BC's, in connection with the problem of breaking
supersymmetry in string theory, was first considered in
Ref.~\cite{Antoniadis:1990ew}. In theories using only usual orbifold BC's \cite{Dixon:jw} for gauge symmetry breaking, however, it is generally difficult to reduce the rank of a gauge group, as it
would be required for realistic EWSB.
Rank reduction, on the other hand,
is easily achieved in the recently proposed new type of Higgsless models for
EWSB \cite{Csaki:2003dt,Csaki:2003zu,Nomura:2003du,Csaki:2003sh,Barbieri:2003pr}, which employ mixed (neither Dirichlet nor Neumann) BC's.\footnote{For GUT
breaking with mixed BC's see Ref.~\cite{Nomura:2001mf}.}
The mixed BC's, when consistent with the variation of a gauge invariant
action, correspond to a soft breaking of the gauge symmetry, since they
can be ultraviolet completed by a boundary Higgs field.

The original model for Higgsless EWSB \cite{Csaki:2003dt} is an
$SU(2)_L\times SU(2)_R\times U(1)_{B-L}$ gauge theory
compactified on an interval $[0,\pi R]$ in five-dimensional (5D) flat space.
At one end of the interval, $SU(2)_R\times U(1)_{B-L}$ is broken to $U(1)_Y$. At the other end, $SU(2)_L\times SU(2)_R$ is broken to the diagonal subgroup
$SU(2)_D$, thereby leaving only $U(1)_Q$ of electromagnetism unbroken in the
effective four-dimensional (4D) theory. Although this
model exhibited some
similarities with the SM, the $\rho$ parameter deviated from unity by
$\sim 10\%$ and the lowest KK excitations of the $W^\pm$ and
$Z$ were too light ($\sim$ 240 GeV) to be in agreement with experiment.
These problems have later been resolved by considering the setup in the
truncated anti de-Sitter (AdS) space of the Randall-Sundrum model
\cite{Randall:1999vf}. Here, the generators broken on the Planck brane can be associated via the AdS/CFT
correspondence \cite{Maldacena:1997re} in the 4D dual
\cite{Arkani-Hamed:2000ds} theory with a global custodial
$SU(2)$ symmetry \cite{Agashe:2003zs}, while the electroweak symmetry
has been broken by the presence of the TeV brane
alone \cite{Csaki:2003zu}. As a consequence,
in the strongly coupled 4D theory, violation of custodial isospin remains (even after inclusion of radiative corrections) only of order
$\sim 1\%$, while the higher KK resonances of the gauge bosons
would decouple below $\sim 1\:{\rm TeV}$ \cite{Csaki:2003zu,Nomura:2003du}.
In this framework, it is possible to generate realistic quark and lepton masses
with viable couplings to $W^\pm$ and $Z$,
when the fermions propagate in the bulk \cite{Nomura:2003du,Csaki:2003sh}. Based on the same gauge group, similar effects can be realized in
5D flat space \cite{Barbieri:2003pr}, when 4D brane kinetic terms
\cite{Mirabelli:1997aj,Georgi:2000ks,Goldberger:2001tn} dominate the
contribution from the bulk. In fact, brane kinetic
terms seem also to be required in Higgsless warped space models
\cite{Davoudiasl:2003me}, to evade disagreement with EWPT due to
tree-level ``oblique'' corrections \cite{Peskin:1991sw,Altarelli:1990zd,Holdom:1990tc}.

In 5D Higgsless models, a $\rho$ parameter close to unity is achieved
at the expense of enlarging the SM gauge group by an additional gauge group
$SU(2)_R$, which introduces a gauged custodial symmetry in the bulk.
Inspired by dimensional deconstruction \cite{Arkani-Hamed:2001ca,Hill:200mu}, 
one can consider the $SU(2)_L\times SU(2)_R$ subgroup of the model as belonging to a chain of 5D
gauge theories with product group structure
$SU(2)_1\times SU(2)_2\times \ldots \times SU(2)_N\supset SU(2)_L\times SU(2)_R$, which is broken down to $SU(2)_D$ by BC's (for a discussion of
Higgsless EWSB in deconstruction see Ref.~\cite{Foadi:2003xa}).
From the deconstruction point of view, such a product group may be reduced
to a single six-dimensional (6D) parent gauge group $SU(2)_L$, while keeping
essential features of the corresponding 5D theory. Hence,
it should be possible to obtain
consistent 6D Higgsless models of EWSB, which are based only on the
SM gauge group $SU(2)_L\times U(1)_Y$ and allow the $\rho$ parameter to be set
equal to unity. There is yet another advantage of going beyond five dimensions.
In more than five dimensions, the physical space can be reduced ({\it e.g.}, by orbifold BC's) to a domain smaller than the periodicity
of the wavefunctions. As a result, the $S,T,$ and $U$ parameters \cite{Peskin:1991sw} would become suppressed by higher powers of the loop expansion parameter of the
theory, thereby potentially improving the calculability of Higgsless models.

In this paper, we consider a Higgsless model for EWSB in six dimensions, which
is based only on the SM gauge group $SU(2)_L\times U(1)_Y$, where the 
gauge bosons propagate in the bulk. The model is formulated in flat space with
the two extra dimensions compactified on a rectangle and
EWSB is achieved by imposing consistent 
BC's. The higher KK
resonances of $W^\pm$ and $Z$ decouple below $\sim 1{\rm TeV}$ through the
presence of a dominant 4D brane induced gauge kinetic term. The
$\rho$ parameter is arbitrary and can be set exactly to one by an appropriate choice of the bulk gauge
couplings and compactification scales. Unlike in the 5D theory, the mass scale of the lightest gauge bosons $W$ and $Z$ is solely set by the dimensionful
bulk couplings, which  (upon compactification via mixed BC's)
are responsible for EWSB. 
We calculate the tree-level oblique corrections to the $S,T,$ and $U$
parameters and find that they are in better agreement with data than in
proposed 5D warped and flat Higgsless models. Non-oblique corrections, however, can generally lead to a tension between the bottom quark mass and the $Z\rightarrow b\overline{b}$ coupling, which could be modified at the level of current experimental uncertainties. By considering the scattering of a scalar propagating in 
$S^1/Z_2$ and $S^1/(Z_2\times Z_2')$ extra dimensions, we estimate the raising
of the strong coupling scale, which
could improve the calculability of Higgsless models formulated on these manifolds.

The paper is organized as follows. In Sec.~\ref{sec:model}, we introduce the 6D model on a rectangle and discuss the symmetry
breaking by BC's. In Sec.~\ref{sec:effective}, we determine the wavefunctions in the presence of the brane terms, vacuum polarizations and KK spectra of the gauge bosons. We compare the oblique corrections to EWPT 
in Sec.~\ref{sec:EWPT}. Non-oblique corrections of the SM couplings
due to the generation of heavy fermion masses are then discussed in Sec.~\ref{sec:fermions}.
Next, in
Sec.~\ref{sec:calculability}, we estimate the strong coupling scale on different orbifold extra dimensions
and outline potential implications for an improved calculability of
Higgsless models. Finally, in Sec.~\ref{sec:summary}, we present our summary
and conclusions.
 
\section{The model}\label{sec:model}
Let us consider a 6D $SU(2)_L\times U(1)_Y$ gauge theory in a flat space-time background, where the two extra spatial dimensions are compactified on a rectangle\footnote{Chiral compactification on a square has recently been considered in Ref.~\cite{Dobrescu:2004zi}.}.The coordinates in the 6D space are written as $z_M=(x_\mu,y_m)$, where the 6D Lorentz indices are denoted by
capital Roman letters $M=0,1,2,3,5,6$, while the usual 4D Lorentz indices are
symbolized by Greek letters $\mu=0,1,2,3$, and the coordinates
$y_m$ $(m=1,2)$ describe the fifth and sixth
dimension.\footnote{For the metric we choose a signature $(+,-,-,-,-,-)$.}
 The physical space is thus defined by $0\leq y_1\leq \pi R_1$ and $0\leq y_2\leq\pi R_2$, where $R_1$ and $R_2$ are the compactification radii of a torus
$T^2$, which is obtained by identifying the points of the two-dimensional plane
$R^2$ under the actions $T_5:(y_1,y_2)\rightarrow(y_1+2\pi R_1,y_2)$
and $T_6:(y_1,y_2)\rightarrow(y_1,y_2+2\pi R_2)$.
We denote the $SU(2)_L$ and $U(1)_Y$ gauge bosons in the bulk respectively by
$A_M^a(z_M)$ ($a=1,2,3$ is the gauge index) and $B_M(z_M)$.
The action of the gauge fields in our model is given by
\begin{equation}\label{eq:S}
 \mathcal{S}=\int d^4x\int_0^{\pi R_1}dy_1\int_0^{\pi R_2}
dy_2\left(\mathcal{L}_6+\delta(y_1)\delta(y_2)\mathcal{L}_{0}\right),
\end{equation}
where $\mathcal{L}_6$ is a 6D bulk gauge kinetic term and
$\mathcal{L}_0$ is a 4D brane gauge kinetic term localized at
$(y_1,y_2)=(0,0)$, which read respectively
\begin{equation}\label{eq:gaugekinetic}
 \mathcal{L}_6  = 
 -\frac{M_L^2}{4}F^a_{MN}F^{MNa}-\frac{M_Y^2}{4}B_{MN}B^{MN},\quad
 \mathcal{L}_0  =  
 -\frac{1}{4g^2}F_{\mu\nu}^a F^{\mu\nu a}-\frac{1}{4{g'}^2}B_{\mu\nu}
B^{\mu\nu},
\end{equation}
with field strengths
$F_{MN}^a=\partial_M A^a_N-\partial_N A^a_M+f^{abc}A^b_MA^c_N$
($f^{abc}$ is the structure
constant) and $B_{MN}=\partial_MB_N-\partial_N B_M$. In Eqs.~(\ref{eq:gaugekinetic}), the quantities $M_L$ and $M_Y$ have mass dimension $+1$, while
$g$ and $g'$ are dimensionless. Since the boundaries of the manifold
break translational invariance and are ``singled out'' with respect to the points in the interior of the rectangle, brane terms like $\mathcal{L}_0$
can be produced by quantum loop effects \cite{Mirabelli:1997aj,Georgi:2000ks}
or arise from classical singularities in the limit of vanishing
brane thickness \cite{Goldberger:2001tn}.

Unlike in five dimensions (for a discussion of the
$\xi\rightarrow\infty$ limit in generalized 5D $R_\xi$ gauges see, {\it e.g.},
Ref.~\cite{Muck:2001yv} and also Ref.~\cite{Csaki:2003dt}), we cannot go to
a unitary gauge where all fields $A^a_{5,6}$
$(a=1,2,3)$ and $B_{5,6}$ are identically set to zero. Instead, there will
remain after dimensional reduction one combination of physical scalar fields
in the spectrum\footnote{We thank H. Murayama and M. Serone for pointing out this fact.}. To make these scalars sufficiently heavier than the
Lee-Quigg-Thacker bound of $\approx 2\:{\rm TeV}$, we can assume, {\it e.g.},
a seventh dimension compactified on $\mathcal{S}^1/Z_2$ with compactification
radius $R_3\lesssim R_1, R_2$. By setting $A_{5,6,7}^a=B_{5,6,7}=0$ ($A^a_{7}$ and $B_7$ are the seventh components of the gauge fields) on all 
boundaries of this manifold, the associated scalars can acquire for compactification scales $R^{-1}_1,R_2^{-1}\simeq 1-2\:{\rm TeV}$, masses well above
$2\:{\rm TeV}$. Therefore, at low energies
$\lesssim 2-3\:{\rm TeV}$, we have a model without any {\it light}
scalars and will, in what follows, neglect the heavy scalar degrees of freedom.

Since the Lagrangian in Eq.~(\ref{eq:gaugekinetic}) does not contain any explicit gauge symmetry breaking, we can obtain consistent new BC's on the boundaries
by requiring the variation of the action to be zero.
Variation of the action in Eq.~(\ref{eq:gaugekinetic}) yields after partial
integration
\begin{eqnarray}\label{eq:variation}
 \delta S &=&\int d^4x\int_{y_1=0}^{\pi R_1}dy_1\int_{y_2=0}^{\pi R_2}dy_2
\left[M_L^2
\left(\partial_MF^{aM\mu}-f^{abc}F^{bM\mu}A^c_M\right)\delta A^a_\mu
+M_Y^2\partial_MB^{M\mu}\delta B_\mu\right]\nonumber\\
&+&\int d^4 x\int_{y_2=0}^{\pi R_2}dy_2
\left[M_L^2 F_{5\mu}^a\delta A^{a\mu}+M_Y^2 B_{5\mu}\delta B^\mu\right]_{y_1=0}^{\pi R_1}\nonumber\\
&+&\int d^4x\int_{y_1=0}^{\pi R_1}dy_1
\left[M_L^2 F_{6\mu}^a\delta A^{a\mu}+M_Y^2 B_{6\mu}\delta B^\mu\right]_{y_2=0}^{\pi R_2}\nonumber\\
&+&\int d^4x\left[\frac{1}{g^2}
(\partial_\mu F^{a\mu\nu}-f^{abc}F^{b\mu\nu}A^c_\mu)\delta A^c_\nu
+\frac{1}{{g'}^2}\partial_\mu B^{\mu\nu}\delta B_\nu\right]_{(y_1,y_2)=(0,0)}\:\:=\:\:0,
\end{eqnarray}
where we have (as usual) assumed that the gauge fields and their derivatives go to zero for $x_\mu\rightarrow \infty$. The bulk terms in
in the first line in Eq.~(\ref{eq:variation}), lead to the familiar bulk
equations of motion. Moreover, since the minimization of the action requires
the boundary terms to vanish as well, we obtain from the second and third line in
Eq.~(\ref{eq:variation}) a set of consistent BC's for the bulk fields.

\begin{figure}
\begin{center}
\includegraphics*[bb = 163 588 449 752]{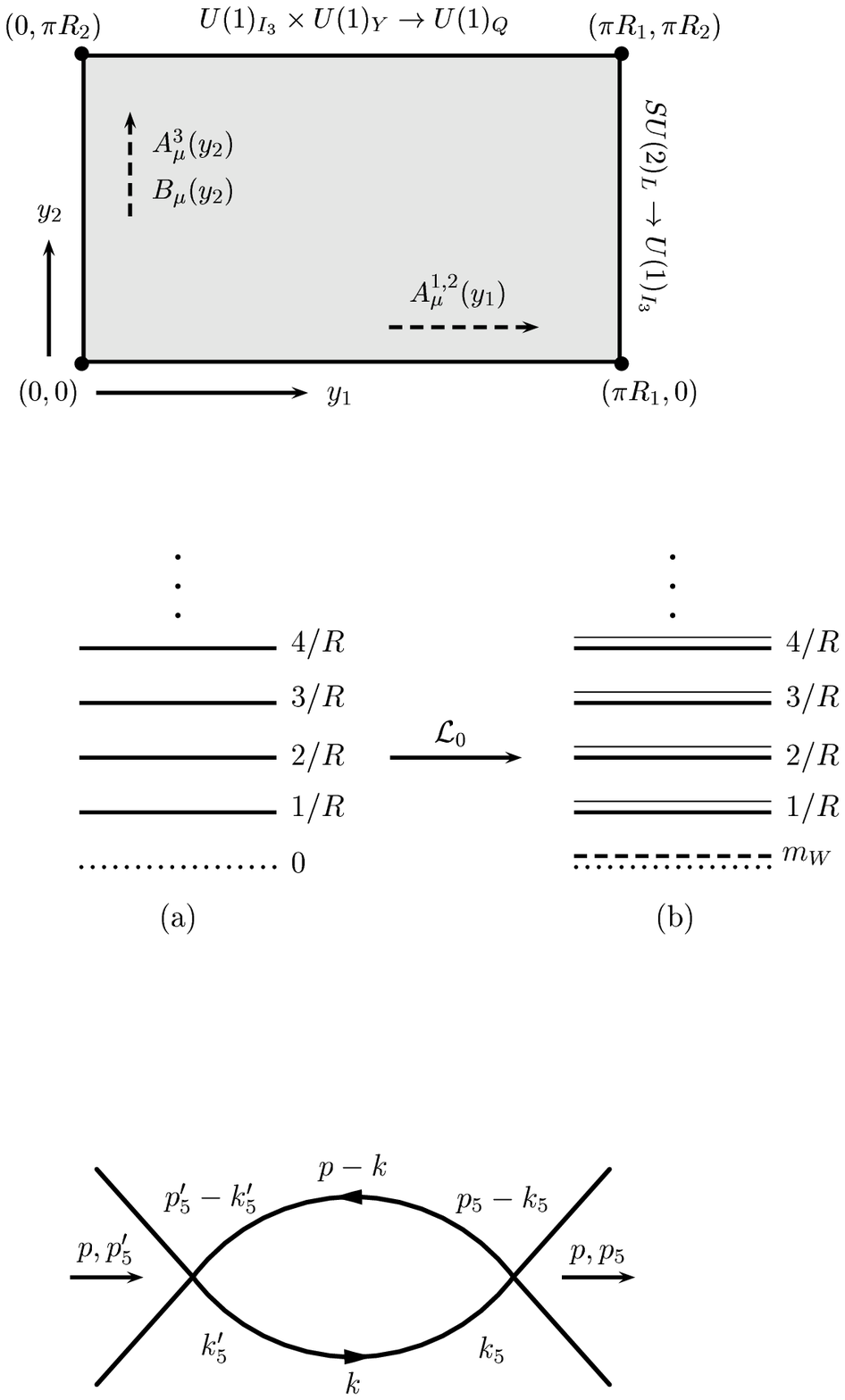}
\caption{{\small{Symmetry breaking of $SU(2)_L\times U(1)_Y$ on the rectangle.
At one boundary $y_1=\pi R_1$}, $SU(2)_L$ is broken to $U(1)_{I_3}$ while
on the boundary $y_2=\pi R_2$ the subgroup
$U(1)_{I_3}\times U(1)_Y$ is broken to $U(1)_Q$, which leaves
only $U(1)_Q$ unbroken on the entire rectangle. Locally, at the fixed
point $(0,0)$, $SU(2)_L\times U(1)_Y$ remains unbroken. The dashed arrows indicate the propagation of the lowest resonances of the gauge bosons.}}\label{fig:rectangle}
\end{center}
\end{figure}

We break the electroweak symmetry $SU(2)_L\times U(1)_Y\rightarrow U(1)_Q$ by imposing on two of the boundaries following BC's:
\begin{subequations}\label{eq:mixed}
\begin{eqnarray}
  {\rm at}\:\:y_1=\pi R_1&:&
  A_\mu^1=0,\quad A_\mu^2=0,\label{eq:mixed1}\\
 {\rm at}\:\:y_2=\pi R_2&:&
 \partial_{y_2}( M_L^2A^3_\mu+M_Y^2B_\mu)=0,\: A^3_\mu-B_\mu=0.
\label{eq:mixed2}
\end{eqnarray}
\end{subequations}
The Dirichlet BC's in Eq.~(\ref{eq:mixed1}) break
$SU(2)_L\rightarrow U(1)_{I_3}$, where $U(1)_{I_3}$ is the $U(1)$ subgroup
associated with the third component of weak isospin $I_3$. The
BC's in Eq.~(\ref{eq:mixed2}) break
$U(1)_{I_3}\times U(1)_Y\rightarrow U(1)_Q$, leaving only $U(1)_Q$ unbroken
on the entire rectangle (see Fig.~\ref{fig:rectangle}). Note, in Eq.~(\ref{eq:mixed2}), that the first BC
involving the derivative with respect to $y_2$ actually follows
from the second BC $\delta A_\mu^3=\delta B_\mu$ by minimization of
the action. The gauge groups $U(1)_{I_3}$ and 
$U(1)_{I_3}\times U(1)_Y$ remain unbroken at the boundaries $y_1=0$ and
$y_2=0$, respectively. Locally, at the fixed point
$(y_1,y_2)=(0,0)$, $SU(2)_L\times U(1)_Y$ is unbroken. We can restrict ourselves, for simplicity, to the solutions which are relevant to EWSB, by imposing on the other two boundaries the following Dirichlet BC's:
\begin{subequations}\label{eq:Dirichlet}
\begin{eqnarray}
  {\rm at}\:\:y_1=0&:&
 A_\mu^{1,2}(z_M)=\overline{A}^{1,2}_\mu(x_\mu),
\label{eq:Dirichlet1}\\
 {\rm at}\:\:y_2=0&:&
 A^3_\mu(z_M)=\overline{A}^3_\mu(x_\mu),\:
 B_\mu(z_M)=\overline{B}_\mu(x_\mu),\label{eq:Dirichlet2}
\end{eqnarray}
\end{subequations}
where the bar indicates a boundary field. The Dirichlet BC's in
Eqs.(\ref{eq:Dirichlet}) require $A^{1,2}_\mu$ to be independent of
$y_2$, while $A^3_\mu$ and $B_\mu$ become independent of $y_1$, such that we
can generally write $A^{1,2}_\mu= A^{1,2}(x_\mu,y_1)$,
$A^3_\mu= A^3_\mu(x_\mu,y_2)$, and $B_\mu= B_\mu(x_\mu,y_2)$.
For the transverse\footnote{Note that $\partial_MF^{aM\mu}=p^2P_{\mu\nu}(p)A^{a\mu}+(\partial^2_{y_1}+\partial^2_{y_2})A^a_\nu=0$, where
$P_{\mu\nu}(p)=g_{\mu\nu}-p_\mu p_\nu/p^2$ is the operator projecting onto transverse states.} components of the gauge fields the bulk equations of motion
then take the forms
\begin{equation}\label{eq:eom}
 (p^2+\partial_{y_1}^2)A^{1,2}_\mu(x_\mu,y_1)=0,\quad
 (p^2+\partial_{y_2}^2)A^3_\mu(x_\mu,y_2)=0,\quad
 (p^2+\partial_{y_2}^2)B_\mu(x_\mu,y_2)=0,
\end{equation}
where $p^2=p_\mu p^\mu$ and $p_\mu=i\partial_\mu$ is the momentum in
the uncompactified 4D space. Since we assume all the gauge couplings to be small,
we will, in what follows, treat
$A_\mu^a$ approximately as a ``free'' field ({\it i.e.}, without self
interaction) and drop all cubic and quartic terms in $A_\mu^a$.

We assume that the fermions, in the first approximation, are localized on the brane at
$(y_1,y_2)=(0,0)$, away from the walls of electroweak symmetry breaking. This choice will avoid any unwanted non-oblique corrections to the electroweak
precision parameters.

\section{Effective theory}\label{sec:effective}
\begin{figure}
\begin{center}
\includegraphics*[bb = 190 382 495 539, height=5cm]{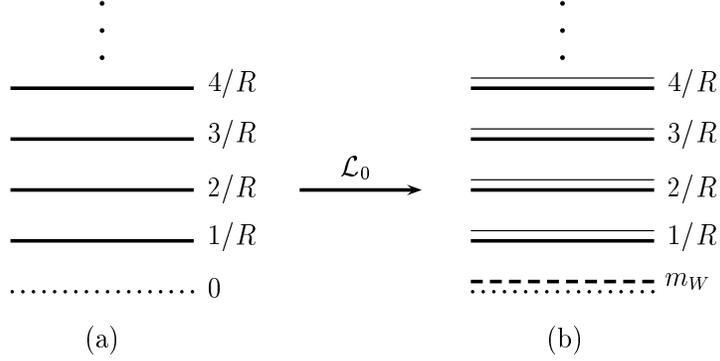}
\vspace*{-2mm}
\caption{\small{Effect of the brane kinetic terms $\mathcal{L}_0$ on the
KK spectrum of the gauge bosons (for the example of $W^\pm$). Solid lines represent massive excitations, the bottom dotted lines would correspond to the zero modes which have been removed by the BC's. Without the
brane terms (a), the lowest KK excitations are of order
$1/R\simeq 1\:{\rm TeV}$. After switching on the dominant brane kinetic
terms (b), the zero modes are approximately ``restored''
with a small mass $m_W\ll 1/R$ (dashed line), while the higher KK-levels
receive small corrections to their masses (thin solid lines) and decouple below
$\sim 1\:{\rm TeV}$.}}\label{fig:braneterms}
\end{center}
\end{figure}

The total effective
4D Lagrangian in the compactified theory $\mathcal{L}_{\rm total}$ can be
written as $\mathcal{L}_{\rm total}=\mathcal{L}_0+\mathcal{L}_{\rm eff}$, where
$\mathcal{L}_{\rm eff}=\int_{0}^{\pi R_1}
dy_1\int_0^{\pi R_2}dy_2\:\mathcal{L}_6$ denotes the contribution from the
bulk, which follows from integrating out the extra dimensions.
After partial integration along the
$y_1$ and $y_2$ directions, we obtain for $\mathcal{L}_{\rm eff}$ the
non-vanishing boundary term
\begin{equation}\label{eq:Leff}
\mathcal{L}_{\rm eff} = -M_L^2\pi R_2
\left[\overline{A}^1_\mu\partial_{y_1}A^{1\mu}+
 \overline{A}^2_\mu\partial_{y_1}A^{2\mu}\right]_{y_1=0}
-\pi R_1\left[M_L^2\overline{A}^3_\mu\partial_{y_2}A^{3\mu}
+M_Y^2\overline{B}_\mu\partial_{y_2}B^\mu\right]_{y_2=0},
\end{equation}
where we have applied the bulk equations of motion and eliminated the terms
from the boundaries at $y_1=\pi R_1$ and $y_2=\pi R_2$ by virtue of the BC's
in Eqs.~(\ref{eq:mixed}). Notice, that in arriving at Eq.~(\ref{eq:Leff}) we
have redefined the bulk gauge fields as
$A_\mu\rightarrow A_\mu'\equiv A_\mu/\sqrt{2}$ to canonically
normalize the kinetic energy terms of the KK modes. In order to determine $\mathcal{L}_{\rm total}$ explicitly, we first solve
the equations of motion in Eq.~(\ref{eq:eom}) and insert the solutions into
the expression for $\mathcal{L}_{\rm eff}$ in Eq.~(\ref{eq:Leff}). The most
general solutions for Eqs.~(\ref{eq:eom}) can be written as
\begin{subequations}
\begin{eqnarray}
 A^{1,2}_\mu(x_\mu,y_1)&=&\overline{A}^{1,2}_\mu(x_\mu)
\:{\rm cos}(py_1)+b_\mu^{1,2}(x_\mu)\:{\rm sin}(py_1),\\
 A^3_\mu(x_\mu,y_2)&=&\overline{A}^3_\mu(x_\mu)
\:{\rm cos}(py_2)+b_\mu^3(x_\mu)\:{\rm sin}(py_2),\\
 B_\mu(x_\mu,y_2)&=&\overline{B}_\mu(x_\mu)
\:{\rm cos}(py_2)+b_\mu^Y(x_\mu)\:{\rm sin}(py_2),
\end{eqnarray}
\end{subequations}
where $p=\sqrt{p_\mu p^\mu}$ and we have already applied the BC's in
Eq.~(\ref{eq:Dirichlet}). The coefficients $b_\mu^a(x_\mu)$ and
$b_\mu^Y(x_\mu)$ are then determined from the BC's in Eqs.~(\ref{eq:mixed}).
For $b_\mu^{1,2}(x_\mu)$, {\it e.g.}, we find from the BC's in
Eq.~(\ref{eq:mixed1}) that
$b^{1,2}_\mu(x_\mu)=-\overline{A_\mu}^{1,2}(x_\mu)\:{\rm cot}(p\pi R_1)$ and hence one obtains
\begin{subequations}\label{eq:wavefunctions}
\begin{equation}\label{eq:A12}
 A^{1,2}_\mu(x_\mu,y_1)=\overline{A}^{1,2}_\mu(x_\mu)
\left[{\rm cos}(py_1)-{\rm cot}(p\pi R_1)\:{\rm sin}(py_1)\right].
\end{equation}
In a similar way, one arrives after some calculation at the solutions
\begin{eqnarray}
A_\mu^3(x_\mu,y_2)&=&
\overline{A}^3_\mu(x_\mu)\left[
{\rm cos}(py_2)+\frac{M_L^2\:{\rm tan}(p\pi R_2)-M_Y^2
\:{\rm cot}(p\pi R_2)}{M_L^2+M_Y^2}\:{\rm sin}(py_2)\right]\nonumber\\
&+&\overline{B}_\mu(x_\mu)
\frac{M_Y^2\:{\rm tan}(p\pi R_2)+M_Y^2\:{\rm cot}(p\pi R_2)}{M_L^2+M_Y^2}\:{\rm sin}(py_2),
\end{eqnarray}
\begin{eqnarray}
B_\mu(x_\mu,y_2)&=&
\overline{A}^3_\mu (x_\mu)
\frac{M_L^2\:{\rm tan}(p\pi R_2)+M_L^2\:{\rm cot}(p\pi R_2)}{M_L^2+M_Y^2}
\:{\rm sin}(p y_2)\nonumber\\
&+&\overline{B}_\mu(x_\mu)\left[{\rm cos}(py_2)+\frac{
M_Y^2\:{\rm tan}(p\pi R_2)-M_L^2\:{\rm cot}(p\pi R_2)}
{M_L^2+M_Y^2}\:{\rm sin}(py_2)
\right].
\end{eqnarray}
\end{subequations}
Inserting the wavefunctions in Eqs.~(\ref{eq:wavefunctions}) into the effective Lagrangian in Eq.~(\ref{eq:Leff}), we can rewrite
$\mathcal{L}_{\rm eff}$ as
\begin{equation}\label{eq:polarizations}
\mathcal{L}_{\rm eff}=\overline{A}^a_\mu\Sigma_{aa}(p^2)\overline{A}^{a\mu}
+\overline{A}^3_\mu\Sigma_{3B}(p^2)\overline{B}^\mu+
\overline{B}_\mu\Sigma_{BB}(p^2)\overline{B}^\mu,
\end{equation}
where $(aa)=(11),(22)$, and $(33)$ and the momentum-dependent coefficients $\Sigma$
are given by
\begin{eqnarray}\label{eq:sigmas}
 \Sigma_{11}(p^2)&=&
\Sigma_{22}(p^2)\:\:=\:\:\pi R_2M_L^2\:p\:{\rm cot}(p\pi R_1),
 \nonumber\\
 \Sigma_{33}(p^2) & =&-\pi R_1M_L^2\:p\:\frac{M_L^2\:{\rm tan}(p\pi R_2)-M_Y^2\:{\rm cot}(p\pi R_2)}{M_L^2+M_Y^2},\nonumber\\
\Sigma_{3B}(p^2)&=&-2\pi R_1 M_L^2M_Y^2p\:\frac{{\rm tan}(p\pi R_2)+{\rm cot}(p\pi R_2)}{M_L^2+M_Y^2},\nonumber\\
\Sigma_{BB}(p^2)&=&-\pi R_1
 M_Y^2p\:\frac{M_Y^2\:{\rm tan}(p\pi R_2)-M_L^2\:{\rm cot}(p\pi R_2)}{M_L^2+M_Y^2}.
\end{eqnarray}
The $\Sigma$'s can be viewed as the electroweak vacuum polarization
amplitudes which summarize in the low energy theory the effect of the
symmetry breaking sector. The presence of these terms leads at tree level to
oblique corrections (as opposed to vertex corrections and box diagrams) of the gauge boson propagators and affects
electroweak precision measurements \cite{Peskin:1991sw,Altarelli:1990zd}. Since
$\mathcal{L}_{\rm eff}$ in Eq.~(\ref{eq:Leff}) generates effective mass
terms for the gauge bosons in the 4D theory\footnote{For an effective field
theory approach to oblique corrections see, {\it e.g.},
Ref.~\cite{Holdom:1990tc}.}, the KK masses of the $W^{\pm}$ bosons are found
from the zeros of the inverse propagator as given by the solutions of the
equation
\begin{equation}\label{eq:Wpropagator}
 \Sigma_{11}(p^2)-\frac{p^2}{2g^2}=0.
\end{equation}
To determine the KK masses of the gauge bosons, we will from now on assume
that the brane terms $\mathcal{L}_0$ dominate the bulk kinetic terms,
{\it i.e.}, we take $1/g^2,1/{g'}^2\gg (M_{L,Y}\pi)^2R_1R_2$. As a result, we
find for the $W^{\pm}$'s the mass spectrum
\begin{eqnarray}\label{eq:mW}
 m_n&=&\frac{n}{R_1}\left(1+\frac{2g^2 M_L^2R_1R_2}{n^2}+\dots\right),
\quad n=1,2,\ldots,\nonumber\\
 m_0^2&=&\frac{2g^2M_L^2R_2}{R_1}+\mathcal{O}(g^4M_L^4R_2^2)
\:\:=\:\:m_W^2,
\end{eqnarray}
where we identify the lightest state with mass $m_0$ with the $W^\pm$. Observe
in Eq.~(\ref{eq:mW}), that the inclusion of the brane kinetic terms
$\mathcal{L}_0$ for $1/R_1,1/R_2 \gtrsim
\mathcal{O}({\rm TeV})$ leads to a decoupling of the higher KK-modes with masses
$m_n$ $(n>0)$ from the electroweak scale, leaving only the $W^\pm$ states
with a small mass $m_0$ in the low-energy theory (see Fig.~\ref{fig:braneterms}). Note that a similar effect has been
found for warped models in Ref.~\cite{Carena:2002dz}.

The calculation of the mass of the $Z$ boson goes along the same lines as for $W^\pm$, but requires, due to the
mixing of $\overline{A}^3_\mu$ with $\overline{B}_\mu$ in Eq.~(\ref{eq:polarizations}), the diagonalization of the kinetic matrix
\begin{equation}
M_{\rm kin}=
\left(
\begin{matrix}
\Sigma_{33}(p^2)-\frac{p^2}{2g^2} & \frac{1}{2}\Sigma_{3B}(p^2)\\
\frac{1}{2}\Sigma_{3B}(p^2) & \Sigma_{BB}(p^2)-\frac{p^2}{2{g'}^2}
\end{matrix}
\right),
\end{equation}
which has the eigenvalues
\begin{eqnarray}\label{eq:eigenvals}
 \lambda_{\pm}(p^2)&=&\frac{1}{2}\left(
\Sigma_{33}(p^2)-\frac{p^2}{2g^2}+\Sigma_{BB}(p^2)-\frac{p^2}{2{g'}^2}
\right)\nonumber\\
&\pm&\frac{1}{2}\sqrt{\left(
\Sigma_{33}(p^2)-\frac{p^2}{2g^2}-\Sigma_{BB}+\frac{p^2}{2{g'}^2}
\right)^2+\Sigma^2_{3B}(p^2)},
\end{eqnarray}
where the KK towers of the $\gamma$ and $Z$ are given by the solutions of the
equations $\lambda_{-}(p^2)=0$ (for $\gamma$) and $\lambda_{+}(p^2)=0$ (for
$Z$), respectively. By taking in Eq.~(\ref{eq:eigenvals}) the limit
$p^2\rightarrow 0$, it is easily seen that $\lambda_{-}(p^2)=0$ has a solution
with $p^2=0$, which we identify with the massless $\gamma$ of the SM,
corresponding to the unbroken gauge group $U(1)_Q$. The lowest excitation in the tower of solutions to $\lambda_{+}(p^2)=0$ has a mass-squared
\begin{equation}\label{eq:mZ}
 m_Z^2=\frac{2(g^2+{g'}^2)M_L^2M_Y^2 R_1}{(M_L^2+M_Y^2)R_2}
+\mathcal{O}(g^4M_L^4R_2^2),
\end{equation}
which we identify with the $Z$ of the SM. All other KK modes of the
$\gamma$ and $Z$ have masses of order $\gtrsim 1/R_2$ and thus decouple for
$1/R_1,1/R_2\gtrsim\mathcal{O}({\rm TeV})$, leaving only
a massless $\gamma$ and a $Z$ with mass $m_Z$ in the low-energy
theory.

\section{Relation to EWPT}\label{sec:EWPT}
One important constraint on any model for EWSB results from the measurement of the $\rho$ parameter, which is experimentally known to satisfy the relation
$\rho=1$ to better than 1\% \cite{Hagiwara:fs}. In our model, we find from
Eqs.~(\ref{eq:mW}) and (\ref{eq:mZ}) a fit of the natural
zeroth-order SM relation for the $\rho$ parameter in terms of
\begin{equation}\label{eq:rho}
\rho\equiv\frac{m_W^2}{m_Z^2\:{\rm cos}^2\:\theta_W}=
\frac{g^2}{g^2+g'^2}\frac{M_L^2+M_Y^2}{M_Y^2}\left(\frac{R_2}{R_1}\right)^2
\frac{1}{{\rm cos}^2\theta_W}=1,
\end{equation}
where $\theta_W\approx 28.8^\circ$ is the Weinberg angle of the SM.
For definiteness, we will choose in the following
the 4D brane couplings $g$ and $g'$ to satisfy the usual SM relation
$g^2/(g^2+{g'}^2)={\rm cos}^2\theta_W\approx 0.77$. Defining
$\rho=1+\Delta\rho$, we then obtain from Eq.~(\ref{eq:rho}) that
$\Delta\rho =0$ if the bulk kinetic couplings and compactification radii
satisfy the relation
\begin{equation}\label{eq:fit}
(M_L^2+M_Y^2)/M_Y^2=R_1^2/R_2^2.
\end{equation}
Although we can thus set $\Delta\rho=0$ by appropriately dialing the gauge
couplings and the size of the extra dimensions, we observe in
Eq.~(\ref{eq:polarizations}) that $\mathcal{L}_{\rm eff}$ 
introduces a manifest breaking of custodial symmetry
(which transforms the three gauge bosons $A^a_\mu$ among themselves)
and will thus contribute to EWPT via oblique corrections to the SM
parameters.\footnote{Note, however, that in the limit $p^2\rightarrow 0$, we
have $\Sigma_{11}=\Sigma_{33}$, which restores custodial symmetry.}

To estimate the effect of the oblique corrections in our model let us consider
in the 4D effective theory a general vacuum polarization tensor
$\Pi^{\mu\nu}_{AB}(p^2)$ between two
gauge fields $A$ and $B$ which can (for canonically normalized fields)
be expanded as
\cite{Holdom:1990tc}
\begin{equation}\label{eq:polarization}
 i\Pi_{\mu\nu}^{AB}(p^2)=ig_Ag_B
\left[\Pi_{AB}^{(0)}+p^2\Pi^{(1)}_{AB}\right]g_{\mu\nu}
 +p_\mu p_\nu\:{\rm terms},
\end{equation}
where $g_A$ and $g_B$ are the couplings corresponding to the gauge fields $A$
and $B$, respectively. After going in $\mathcal{L}_{\rm eff}$ back to
canonical normalization by redefining
$A_\mu^a\rightarrow A_\mu'\equiv A_\mu^a/g$ and
$B_\mu\rightarrow B_\mu'\equiv B_\mu/g'$, we identify
$\Sigma_{aa}(p^2)\simeq\frac{1}{2}[\Pi^{(0)}_{aa}+p^2\Pi_{aa}^{(1)}]$,
for $(aa)=(11),(22),(33),(BB)$, while
$\Sigma_{3B}(p^2)\simeq\Pi^{(0)}_{3B}+p^2\Pi^{(1)}_{3B}$. From
Eqs.~(\ref{eq:sigmas}) we then obtain the polarization amplitudes
\begin{eqnarray}\label{eq:amplitudes}
\Pi^{(0)}_{11}&=&\Pi^{(0)}_{22}\:=\:2M_L^2\frac{R_2}{R_1},\quad
 \Pi^{(1)}_{11}\:=\:\Pi^{(1)}_{22}\:=\:-2\frac{\pi^2M_L^2}{3}R_1R_2,
 \nonumber\\
\Pi^{(0)}_{33}&=&2\frac{M_L^2M_Y^2}{M_L^2+M_Y^2}\frac{R_1}{R_2},\quad
 \Pi^{(1)}_{33}\:=\:-2\frac{\pi^2M_L^2R_1R_2}{M_L^2+M_Y^2}(M_L^2+\frac{1}{3}
M_Y^2),\nonumber\\
\Pi^{(0)}_{3B}&=&-2\frac{M_L^2M_Y^2}{M_L^2+M_Y^2}\frac{R_1}{R_2},\quad
\Pi^{(1)}_{3B}\:=\:-\frac{4}{3}\frac{\pi^2M_L^2M_Y^2}{M_L^2+M_Y^2}R_1R_2.
\end{eqnarray}
A wide range of effects from new physics on EWPT can be parameterized
in the $\epsilon_1$, $\epsilon_2$, and $\epsilon_3$ framework
\cite{Altarelli:1990zd}, which is related to the $S,T$, and $U$ formalism of
Ref.~\cite{Peskin:1991sw} by $\epsilon_1=\alpha T$,
$\epsilon_2=-\alpha U/4\:{\rm sin}^2\theta_W$, and
$\epsilon_3=\alpha S/4\:{\rm sin}^2\theta_W$. The experimental bounds on the
relative shifts with respect to the SM expectations are roughly of the order
$\epsilon_1,\epsilon_2,\epsilon_3\lesssim 3\cdot 10^{-3}$ \cite{Barbieri:2004qk}. From
Eq.~(\ref{eq:amplitudes}) we then obtain for these parameters explicitly
\begin{subequations}
\begin{eqnarray}
 \epsilon_1&=&g^2(\Pi^{(0)}_{11}-\Pi^{(0)}_{33})/m_W^2
\:=\:
-2g^2\frac{M_L^2}{m_W^2}\frac{R_1}{R_2}\left(M_Y^2/(M_L^2+M_Y^2)-(R_2/R_1)^2\right),\label{eq:epsilon1}\\
\epsilon_2&=&g^2(\Pi^{(1)}_{33}-\Pi^{(1)}_{11})\:=\:-g^2
\frac{4\pi^2}{3}\frac{M_L^4}{M_L^2+M_Y^2}R_1R_2,\label{eq:epsilon2}\\
 \epsilon_3&=&-g^2\Pi^{(1)}_{3B}\:=\:
g^2\frac{4\pi^2}{3}\frac{M_L^2M_Y^2}{M_L^2+M_Y^2}R_1R_2,\label{eq:epsilon3}
\end{eqnarray}
\end{subequations}
where we have used in the last equation that
$-\epsilon_3/(gg')=\Pi^{(1)}_{3\gamma}/{\rm sin}^2\theta_W-\Pi^{(1)}_{33}
= {\rm cot}\:\theta_W\Pi^{(1)}_{3B}$ \cite{Altarelli:1990zd}. Note in
Eq.~(\ref{eq:epsilon1}), that for our choice of parameters
we have $\epsilon_1=\Delta\rho=0$. The quantities $|\epsilon_2|$ and
$|\epsilon_3|$, on the other hand, are bounded from below by the requirement of
having sufficiently many KK modes below the strong coupling (or cutoff)
scale of the theory. Using ``naive dimensional analysis'' (NDA)
\cite{Manohar:1983md,Georgi:1986kr}, one obtains for the strong coupling scale
$\Lambda$ of a $D$-dimensional gauge theory \cite{Chacko:1999hg} roughly
$\Lambda^{D-4}\simeq (4\pi)^{D/2}\Gamma(D/2)/g_D^2$,
where $g_D$ is the bulk gauge coupling. In our 6D model, we would therefore
have $\Lambda\simeq\sqrt{2}(4\pi)^{3/2}M_{L,Y}$ which leads for
$M_{L,Y}\simeq 10^{2}{\rm GeV}$ to a cutoff $\Lambda\simeq 6\:{\rm TeV}$. Assuming for simplicity $M_L=M_Y$, it follows from Eq.~(\ref{eq:fit}) that $R_2=R_1/\sqrt{2}$, and using Eqs.~(\ref{eq:epsilon2}) and
(\ref{eq:epsilon3}) we obtain
\begin{eqnarray}\label{eq:epsilon3value}
 \epsilon_3\simeq\frac{g^2}{96\sqrt{2}\pi}(\Lambda R_2)^2\simeq
  2.3 \times10^{-3}
\times (g\Lambda R_2)^2,
\end{eqnarray}
while $\epsilon_2\simeq\epsilon_3$. It is instructive to compare the value
for $\epsilon_3$ in our 6D setup as given by Eq.~(\ref{eq:epsilon3value}) with the corresponding result of the 5D model in Ref.~\cite{Barbieri:2003pr}. We find that by going from 5D to 6D, the
strong coupling scale of the theory is lowered from $\sim 10\:{\rm TeV}$ down
to $\sim 6\:{\rm TeV}$. Despite the lowering of the cutoff scale, however, the
parameter $\epsilon_3$ is in the 6D model by $\sim 15\%$ smaller than the
corresponding 5D value\footnote{Notice that in Ref.~\cite{Barbieri:2003pr}, the strong
coupling scale is defined by $1/\Lambda=1/\Lambda_L+1/\Lambda_R$, while we 
assume for $M_L=M_Y$ that $\Lambda=\Lambda_L=\Lambda_Y$.}.
This is due to the fact that in the 6D model the bulk gauge kinetic couplings satisfy
$M_L=M_Y\simeq 100\:{\rm GeV}$, while they take in 5D the values
$M_L\simeq M_Y\simeq 10\:{\rm GeV}$, which is one order of magnitude below the
electroweak scale. From
Eq.~(\ref{eq:epsilon3value}) we then conclude that one can take
for the inverse loop expansion parameter $\Lambda R_2\simeq1/g\approx 1.6$ in agreement
with EWPT. Like in the 5D case, however, the 6D model seems not to
admit a loop expansion parameter in the regime $\Lambda R_2\gg 1$ as required
for the model to be calculable.

\section{Non-oblique corrections and fermion masses}\label{sec:fermions}
In the previous discussion, we have assumed that the fermions are (approximately) localized at
$(y_1,y_2)=(0,0)$.  This would make the fermions exactly massless, since they have no access to the EWSB at $y_1=\pi R_1$ and $y_2=\pi R_2$. In this limiting case, the effects on the electroweak precision parameters
$(\epsilon_1,\epsilon_2,\epsilon_3/S,T,U)$ come from the oblique corrections
due to the vector self energies as given by Eq.~(\ref{eq:polarizations}). A more realistic case will be to extend the fermion wave functions to the bulk,
{\it i.e.}, to the walls of EWSB, where fermion mass operators of the form
$C\overline{\Psi}_L\Psi_R$ ($C$ is some appropriate mass parameter)
can be written. Thus, although the fermion wave functions will be dominantly localized at $(0,0)$, the profile of the wavefunctions
in the bulk will be such that it will have small contributions from the symmetry breaking walls, giving rise to fermion masses. The hierarchy of fermion masses would then be accommodated by some suitable choice of the parameters $C$
\cite{Agashe:2003zs}.


To make the incorporation of heavy fermions in our model explicit,
let us introduce the 6D chiral quark fields
$\mathcal{Q}_i$, $\mathcal{U}_i$, and $\mathcal{D}_i$ ($i=1,2,3$ is the
generation index), where $\mathcal{Q}_i$ are the isodoublet quarks,
while $\mathcal{U}_i$ and $\mathcal{D}_i$ denote the isosinglet up and down
quarks, respectively. For the cancellation of the
$SU(3)_C\times SU(2)_L\times U(1)_Y$ gauge and gravitational anomalies we assume that $\mathcal{Q}_i$ have positive and $\mathcal{U}_i,\mathcal{D}_i$ have negative $SO(1,5)$ chiralities
\cite{Dobrescu:2001ae}.
Next, we
consider the action of the top quark fields with zero bulk mass, which is
given by
\begin{eqnarray}\label{eq:fermionaction}
 \mathcal{S}_{\rm fermion}&=&\int dx^4\int_{0}^{\pi R_1}dy_1\int_0^{\pi R_2}dy_2\:
i(\overline{\mathcal{Q}}_3\Gamma^M D_M\mathcal{Q}_3+
\overline{\mathcal{U}}_3\Gamma^MD_M\mathcal{U}_3)\nonumber\\
&+&\int dx^4\int_{0}^{\pi R_1}dy_1\int_0^{\pi R_2}dy_2\:
K\delta(y_1)\delta(y_2)i[\overline{\mathcal{Q}}_3\Gamma^\mu D_\mu\mathcal{Q}_3+
\overline{\mathcal{U}}_3\Gamma^\mu D_\mu\mathcal{U}_3]\nonumber\\
&+&\int dx^4\int_{0}^{\pi R_1}dy_1\int_0^{\pi R_2}dy_2\:
C\delta(y_1-\pi R_1)\delta(y_2-\pi R_2)\overline{\mathcal{Q}}_{3L}\mathcal{U}_{3R}+{\rm h.c.},
\end{eqnarray}
where we have added in the second line 4D brane kinetic terms with a (common)
gauge kinetic parameter $K=[m]^{-2}$ at $(y_1,y_2)=(0,0)$ and in the third
line we included a boundary mass term with coefficient $C=[m]^{-1}$, which mixes
$\mathcal{Q}_{3L}$ and $\mathcal{U}_{3R}$ at $(y_1,y_2)=(\pi R_1,\pi R_2)$. Note, that the
addition of the
boundary mass term in the last line of Eq.~(\ref{eq:fermionaction}) is consistent with gauge invariance, since $U(1)_Q$ the only gauge group surviving at
$(y_1,y_2)=(\pi R_1,\pi R_2)$. Consider now first the limit of a vanishing brane kinetic term $K\rightarrow 0$. Like in the 5D case \cite{Csaki:2003sh}, appropriate Dirichlet and Neumann BC's for $\mathcal{Q}_{3L,R}$ and $\mathcal{U}_{3L,R}$ would give, in the KK tower corresponding to the top quark, a lowest mass
eigenstate, which is a Dirac fermion with
mass $m_t$ of the order $m_t \sim C/R^2$, where we have defined the length
scale $R\sim R_1\sim R_2$. Next, by analogy with the generation of the $W^\pm$ and $Z$ masses, switching on a dominant brane kinetic term $K/R^2\gg 1$, ensures an approximate
localization of $\mathcal{Q}_{3L}$ and $\mathcal{U}_{3R}$ at $(y_1,y_2)=(0,0)$
and leads to $m_t\sim C/K$ \cite{Barbieri:2003pr}. Now,
the typical values of non-oblique corrections to the SM gauge couplings
coming from the bulk are\footnote{The factor $C$ becomes obvious when treating
 the brane fields in Eq.~(\ref{eq:fermionaction}) as 4D fields, in which case $C=[m]^{+1}$ and $K=[m]^0$.}
$\sim CR/K\sim m_t/(1/R)$ and keeping these
contributions under control, the compactification scale $1/R$ must be
sufficiently large. Like in 5D models, this generally introduces a
possible tension between the 3rd generation quark masses and the coupling of
the $Z$ to the bottom quark. Replacing in the above discussion $\mathcal{U}_{3L,R}$ with $\mathcal{D}_{3L,R}$ and $m_t$ by the bottom quark mass $m_b(m_Z)\approx 3\:{\rm GeV}$, we thus estimate for $1/R\sim 1\:{\rm TeV}$ a
shift of the SM $Z\rightarrow\overline{b}_Lb_L$ coupling by roughly $\sim 0.3\%$, which is of the order of current experimental uncertainties\footnote{The LEP/SLC fit of $\Gamma_b/\Gamma_{\rm had}$ in $Z$ decay requires the shift of the $Z\rightarrow \overline{b}_Lb_L$ coupling to be $\lesssim 0.3\%$ \cite{LEP}.}. Similarly, we predict in our model the coupling of the $Z$ to the top quark to deviate by $\sim 10\%$ from the SM value, which can be checked in the electroweak production of single top in the Tevatron Run 2. It can also be tested in the $t\overline{t}$ pair production in a possible future linear collider.

\section{Improving the calculability}\label{sec:calculability}
\begin{figure}
\begin{center}
\includegraphics*[bb = 188 207 415 307,height=3.3cm]{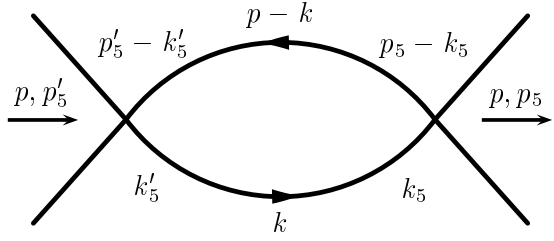}
\caption{{\small{One-loop diagram for $\phi$-$\phi$ scattering on $S^1/Z_2$. The total incoming momentum is $(p,p_5')$ and the total outgoing momentum is
$(p,p_5)$. Generally, it is possible that $|p_5'|\neq |p_5|$, since the orbifold fixed points break 5D translational invariance.}}}
\label{fig:nda}
\end{center}
\end{figure}

To improve the calculability of the model, it seems
necessary to raise (for given $1/g_D^2$) the strong coupling scale
$\Lambda$, which would allow the appearance of more KK modes below the
cutoff. In fact, it has recently been
argued that the compactification of a 5D gauge theory on an orbifold
$S^1/Z_2$ gives a cutoff which is by a factor of 2 larger than the NDA
estimate obtained for an uncompactified space \cite{Barbieri:2004qk}.
Let us now demonstrate this effect
explicitly by repeating the NDA calculation of Ref.~\cite{Manohar:1983md} on an
orbifold following the methods of Refs.~\cite{Georgi:2000ks} and
\cite{Cheng:2002iz}. For this purpose, consider a 5D scalar field
$\phi(x_\mu,y)$ (where we have defined $y=y_1$), propagating in an
$S^1/Z_2$ orbifold extra dimension. The radius of the 5th dimension is $R$ and periodicity
implies $y+2\pi R\sim y$. As a consequence, the momentum in the fifth
dimension is quantized as $p_5=n/R$ for integer $n$. Under the $Z_2$ action
$y\rightarrow -y$ the scalar transforms as
$\phi(x_\mu,y)=\pm\phi(x_\mu,-y)$, where the $+$ $(-)$ sign corresponds to
$\phi$ being even (odd) under $Z_2$. The scalar propagator on this space
is given by \cite{Georgi:2000ks,Cheng:2002iz}
\begin{equation}
  D(p,p_5,p_5')=\frac{i}{2}\left\{
\frac{\delta_{p_5,p_5'}\pm \delta_{-p_5,p_5'}}{p^2-p_5^2}
\right\},
\end{equation}
where the additional factor $1/2$ takes into account that the physical space
is only half of the periodicity.
Consider now the one-loop $\phi$-$\phi$ scattering diagram in
Fig.~\ref{fig:nda}.
The total incoming
momentum is $(p,p_5')$ and the total outgoing momentum is $(p,p_5)$, which can in general be different, since 5D translation invariance is broken by the orbifold
boundaries. Locally, however, momentum is conserved at the vertices.
The diagram then reads
\begin{equation}\label{eq:diagram}
 i\Sigma= \frac{1}{4}\frac{\lambda^2}{2}\frac{1}{2\pi R}\sum_{k_5,k_5'}
\int\frac{d^4k}{(2\pi)^4}\left\{
\frac{\delta_{k_5,k_5'}\pm \delta_{-k_5,k_5'}}{k^2-k_5^2}
\right\}\left\{
\frac{\delta_{(p_5-k_5),(p_5'-k_5')}\pm\delta_{-(p_5-k_5),(p_5'-k_5')}}{(p-k)^2
-(p_5-k_5)^2}\right\},
\end{equation}
where $\lambda$ is the quartic coupling and the additional
factor $1/4$ results from working on $S^1/Z_2$. After
summing over $k_5'$, the integrand can be written as
\begin{equation}\label{eq:integrand}
 F(k_5)=\frac{1}{(k^2-k_5^2)\left[(p-k)^2-(p_5-k_5)^2\right]}
\left\{\delta_{p_5p_5'}+\delta_{p_5,-p_5'}\pm\delta_{2k_5,(p_5+p_5')}\pm
\delta_{2k_5,(p_5-p_5')}\right\}.
\end{equation}
In Eq.~(\ref{eq:integrand}), the first two terms in the bracket conserve
$|p_5'|$ and contribute to the bulk kinetic terms of the scalar. The last two terms, on the other hand, violate $|p_5'|$ conservation and thus lead to a
renormalization of the brane couplings \cite{Georgi:2000ks}. Note that these
brane terms lead in Eq.~(\ref{eq:diagram}) to a logarithmic divergence. 
Applying, on the other hand, to the bulk terms the Poisson resummation identity
\begin{equation}
\frac{1}{2\pi R}\sum_{m=-\infty}^\infty
F(m/R)=\sum_{n=-\infty}^\infty\int_{-\infty}^\infty\frac{dk}{2\pi}
e^{-2\pi ikRn}F(k),
\end{equation}
we obtain a sum of momentum space integrals, where
the ``local'' $n=0$ term diverges linearly like in 5D uncompactified space.
This term contributes a linear divergence to the diagram such that
the scattering amplitude becomes under order one rescalings of the 
random renormalization point for the external momenta of the order
\begin{equation}\label{eq:cutoff}
i\Sigma\rightarrow \frac{\lambda^2}{4}\int\frac{d^5k}{(2\pi)^5}
[k^2(p-k)^2]^{-1}\simeq\frac{\lambda^2}{2}
\frac{\Lambda}{(4\pi)^{5/2}\Gamma(5/2)},
\end{equation}
where $\Lambda$ is an ultraviolet cutoff. On $S^1/Z_2$, we thus indeed
obtain for the strong coupling scale
$\Lambda\simeq 48\pi^3\lambda^{-2}$, which is two times larger than the
NDA value obtained in 5D uncompactified space. This is also in agreement with
the definition of $\Lambda$ for a 5D gauge theory on an interval given in Ref.~\cite{Barbieri:2004qk}.

Similarly, when the 5th dimension is compactified on
$S^1/(Z_2\times Z_2')$ \cite{Barbieri:2000vh}, we expect a raising of
$\Lambda$ by a factor of 4
with respect to the uncompactified case. Let us briefly estimate how far
this could improve the calculability of our 6D model. To this end, we assume,
besides the two extra dimensions compactified on the rectangle, two additional
extra dimensions with radii $R_3$ and $R_4$, each of which
has been compactified on $S^1/(Z_2\times Z_2')$. We assume that the gauge
bosons are even under the actions of the $Z_2\times Z_2'$ groups.
Moreover, we take for the bulk kinetic coefficients in eight dimensions
 $M_L^4=M_Y^4$ and set $R_3=R_4=R_2=R_1/\sqrt{2}$. From the expression analogous to Eq.~(\ref{eq:epsilon3}), we then obtain the estimate
$\epsilon_3\simeq g^2(\pi M_LR_2)^4/3\sqrt{2}$, where the relative factor
$(\pi R_2/2)^2$, arises from integrating over the physical space on each
circle, which is only $1/4$ of the circumference. With respect to the NDA
value $\Lambda^4\simeq (4\pi)^4\Gamma(4)M_L^4$ in uncompactified space, the cutoff gets now modified as
$\Lambda^4\rightarrow 16\cdot\Lambda^4$, implying that
\begin{equation}
 \epsilon_3\simeq \frac{g^2}{192\sqrt{2}}(\Lambda R_2/4)^4
 \simeq 1.3\times 10^{-3}\times(\Lambda R_2/4)^4.
\end{equation}
In agreement with EWPT, the loop expansion parameter could therefore assume
here a value $(\Lambda R_2)^{-1}\simeq 0.25$, corresponding to the
appearance of 4 KK modes per extra dimension below the cutoff. Taking also
a possible additional raising of $\Lambda$ by a factor of $\sqrt{2}$ due
to the reduced physical space on the rectangle into account, one could have
$(\Lambda R_2)^{-1}\simeq 0.2$ with $5$ KK modes per extra dimension below the cutoff. In conclusion, this demonstrates that by going beyond five dimensions, the calculability of Higgsless models could be improved by factors related to
the geometry. 

\section{Summary and conclusions}\label{sec:summary}
In this paper, we have considered a 6D Higgsless model for EWSB
based only on the SM gauge group $SU(2)_L\times U(1)_Y$.
The model is formulated in flat space with
the two extra dimensions compactified on a rectangle of size $\sim ({\rm TeV})^{-2}$. EWSB is achieved by imposing (in the unitary gauge) consistent 
BC's on the edges of the rectangle. The higher KK
resonances of $W^\pm$ and $Z$ decouple below $\sim 1{\rm TeV}$ through the
presence of a dominant 4D brane induced gauge kinetic term at the point where
$SU(2)_L\times U(1)_Y$ remains unbroken. The $\rho$ parameter is arbitrary and can be set exactly to unity by appropriately choosing the bulk gauge
couplings and compactification scales. As a consequence of integrating out
{\it two} extra dimensions, the mass scale of the gauge bosons is essentially independent of the compactification scales and thus set by the bulk gauge
kinetic parameters $M_L$ and $M_Y$ alone, which are of the order of the electroweak scale. The resulting gauge couplings in the effective 4D theory arise essentially from the brane couplings, slightly modified (at the level of one percent) by the bulk interaction. Thus, the main r\^ole played by the bulk interactions is to break the electroweak
gauge symmetry.
We calculate the tree-level oblique corrections to the $S,T,$ and $U$
parameters and find them to be consistent with current data. Non-oblique corrections to the SM gauge couplings, however, can generally modify the coupling of
the $Z$ to the bottom quark at the level of current experimental uncertainties.
By considering at one-loop
the $\phi^4$ interaction of a scalar $\phi$ propagating on $S^1/Z_2$ and
$S^1/(Z_2\times Z_2')$, we estimate the shift of the strong coupling scale for
models formulated on these manifolds. We thus conclude that a stronger
suppression of the tree-level oblique corrections could be obtained in the
presence of one or two extra dimensions (in addition to the ones compactified
on the rectangle), each of which has been compactified on 
$S^1/{(Z_2\times Z_2')}$, thereby improving the  calculability of the model.

\section*{Acknowledgments}
We would like to thank K. Agashe, Ts. Enkhbat, Y. Nomura, and L. Randall for useful comments and discussions. We would also like to thank the theory group at Fermilab for their kind hospitality, where part of this work was done.
This work is supported in part by the Department of Energy under
Grant Nos. DE-FG02-04ER46140 and DE-FG02-04ER41306.

\end{document}